\documentclass[conference]{IEEEtran}
\usepackage{amsmath,amssymb,amsfonts}
\usepackage{algorithm}
\usepackage{algpseudocode}
\usepackage{graphicx}
\usepackage{textcomp}
\usepackage{bbm}
\usepackage{wrapfig}
\usepackage{multicol}
\usepackage{xcolor}
\def\BibTeX{{\rm B\kern-.05em{\sc i\kern-.025em b}\kern-.08em
    T\kern-.1667em\lower.7ex\hbox{E}\kern-.125emX}}
\begin{document}

\title{Universal Fourier Attack for Time Series}

\makeatletter
\newcommand{\linebreakand}{%
  \end{@IEEEauthorhalign}
  \hfill\mbox{}\par
  \mbox{}\hfill\begin{@IEEEauthorhalign}
}
\makeatother


\author{Elizabeth Coda, Brad Clymer, Chance DeSmet, Yijing Watkins, Michael Girard \\
Pacific Northwest National Lab\\
\texttt{\{first.last\}@pnnl.gov} \\
}


\maketitle

\begin{abstract}
A wide variety of adversarial attacks have been proposed and explored using image and audio data. These attacks are notoriously easy to generate digitally when the attacker can directly manipulate the input to a model, but are much more difficult to implement in the real-world. In this paper we present a universal, time invariant attack for general time series data such that the attack has a frequency spectrum primarily composed of the frequencies present in the original data. The universality of the attack makes it fast and easy to implement as no computation is required to add it to an input, while time invariance is useful for real-world deployment. Additionally, the frequency constraint ensures the attack can withstand filtering. We demonstrate the effectiveness of the attack in two different domains, speech recognition and unintended radiated emission, and show that the attack is robust against common transform-and-compare defense pipelines. 
\end{abstract}


\section{Introduction}
The quantity of proposed adversarial attacks for both image and audio data is vast. Generally, these attacks are easy to create and deploy when the attacker can directly modify an image or audio recording that a model receives as an input. However, implementation is more difficult in the real-world where an attacker must interfere with the data as it is collected. In the image domain, this may require printing out a patch or other object and placing it in the scene before the scene is photographed and in the audio domain this may require broadcasting an attack over-the-air while the data is recorded \cite{brown2017, wu2019}.

In addition to the added cost of physically implementing these attacks, real-world attacks are also constrained by physical limitations. For example, several speech attacks implemented digitally propose computing the attack based on a signal and then mixing it back into the signal \cite{carlini2018}. However, with real-time streaming speech data this is infeasible because the attack cannot be calculated until the signal is recorded, and thus the attack cannot be mixed into the signal during recording \cite{chiquier2022, mathov2020}. Moreover, the frequency spectrum of a real-world speech attack is limited by equipment, as many speakers and recording devices are often constrained to emit and record frequencies within the range of human hearing, and is also limited from a defense perspective, as an attack composed of frequencies outside the frequency spectrum of the original, unperturbed data can be removed through filtering. Finally, a speech attack must also be robust against environmental effects such as noise and reverberation \cite{yakura2019}. 

We propose learning a universal, time invariant attack, $v$, for general time-series data such that the frequency spectrum of $v$ matches the frequency spectrum of the original, unperturbed data. Given a trained model $f$, a universal adversarial attack is a single $v$ such that $f(x + v)$ fools the model for most inputs $x$ \cite{moosavi2017}. The universality of the attack does not require us to know the specific signal we are going to attack ahead of time and allows us to efficiently add the attack to a signal. The time invariance of the attack means that we can play the attack on a loop and the effectiveness of the attack will not be sensitive to the alignment of the start of the attack and signal. Finally, the frequency constraint ensures that our attack is robust against basic filtering defenses. We demonstrate that this attack is effective on both speech data and unintended radiated emission data.  



\section{Methods}

\begin{figure*}[h!]
    \centering
    \includegraphics[scale=0.45]{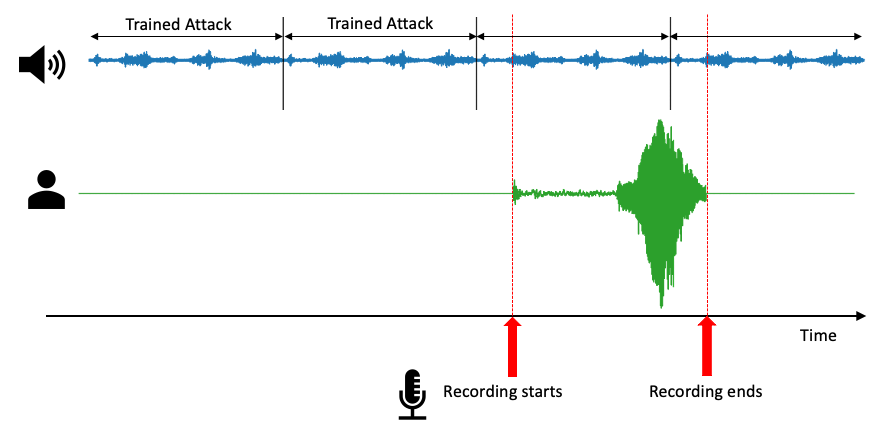}
   \caption{In the proposed implementation, the attack is played repeatedly in the background. In the context of speech recognition, the time when a person starts speaking and the recording begins may not align with the beginning of the attack cycle. We design the attack so that it is time invariant, meaning that the attack remains effective regardless of what point in its cycle the recording begins at. In the URE context, the person can be replaced by a device emitting an electronic signal.}
   \label{fig:looped_attack}
\end{figure*}

\subsection{Data}

\paragraph{Speech Commands}

The Speech Commands dataset is an audio dataset consisting of one-second clips of one-word commands such as 'stop' or 'go' sampled at a rate of 16 kHz \cite{warden2018}. For simplicity, we have removed audio clips labeled as background noise or unknown from the dataset resulting in ten classes with 30k training examples and 3.7k validation examples.

\paragraph{Corona Duff}
The Corona Duff dataset consists of unintended radiated emission (URE) data from 20 common household devices, including a desktop monitor, alarm clock, and a table fan, collected in a residential environment \cite{vann2018}. Voltage and current data were collected from each device over four non-consecutive ten minute runs at a sample rate of 192 kHz. Our training dataset consists of 10k randomly selected 0.1 second segments of voltage data. The validation data consists of 2k randomly selected 0.1 second segments of voltage data, selected from different data collection runs than the training data. Fig.~\ref{fig:learned_examples} includes a visualization of Corona Duff data.

\paragraph{Preprocessing}
For both datasets, we convert the time series to a spectrogram as a preliminary step in the model pipeline. We adjust the length of the FFT used and the step size between FFT windows for each dataset and stack the real and imaginary channels so that the resulting real-valued spectrogram has dimensions 2 x 224 x 224. 

\subsection{Models}
We primarily focus on attacking classifier models. For each of our datasets, we finetune a ResNet18~\cite{he2015} that has been pretrained on ImageNet~\cite{imagenet}. The spectrogram obtained as described above is the input to the classifier. During training, we add Gaussian noise to the signal in the time domain before it is converted to a spectrogram, and then apply random time and frequency masking to the spectrogram. 

\subsection{Metrics}
We use the adversarial success rate (ASR) as our primary evaluation metric. The ASR of an attack is defined as the percentage of originally correct model predictions that the attack successfully changes the prediction of. Unlike the error rate, the ASR only counts inputs where the attack changes the model prediction and does not give the attack credit for inputs the model was originally wrong on. An ASR close to one indicates a highly effective attack. More formally, for a model $f$, attack $v$, and dataset $\mathcal{D} = \{(x_i, y_i) \}_{i=0} ^n$ with inputs $x_i$ and labels $y_i$: 

\begin{equation}
\label{eq:adv_success_rate}
 \text{ASR} = \frac{ \sum_{i=0}^n { \mathbbm{1}{ (f(x_i) = y_i \wedge f(x_i + v) \neq y_i ) } }}{\sum_{i=0}^n {  \mathbbm{1}{ (f(x_i) = y_i)}}}
\end{equation} 

We find the ASR as a function of the signal-to-noise ratio (SNR). As in \cite{yakura2019}, the SNR is $ 10 \text{log}_{10} \frac{P_x}{P_v}$ where $P_x$ is the power of the unperturbed input, $\frac{1}{T} \sum_i^T x_i^2$, and $P_v$ is the power of the attack, $\frac{1}{T} \sum_i^T v_i^2$. The SNR is large when the attack is small and presumably less perceptible. 

Additionally, we compare our models against two simple, baseline adversarial attacks: Fast Gradient Sign Method (FGSM)~\cite{goodfellow2014} and Universal Adversarial Perturbation (UAP)~\cite{moosavi2017}. We emphasize that unlike our attack and the UAP attack, the FGSM attack is not a universal attack, and rather a separate attack is generated for each input to the model. 

\section{Attack}

\begin{figure}[h!]
    \centering
    \includegraphics[scale=0.35]{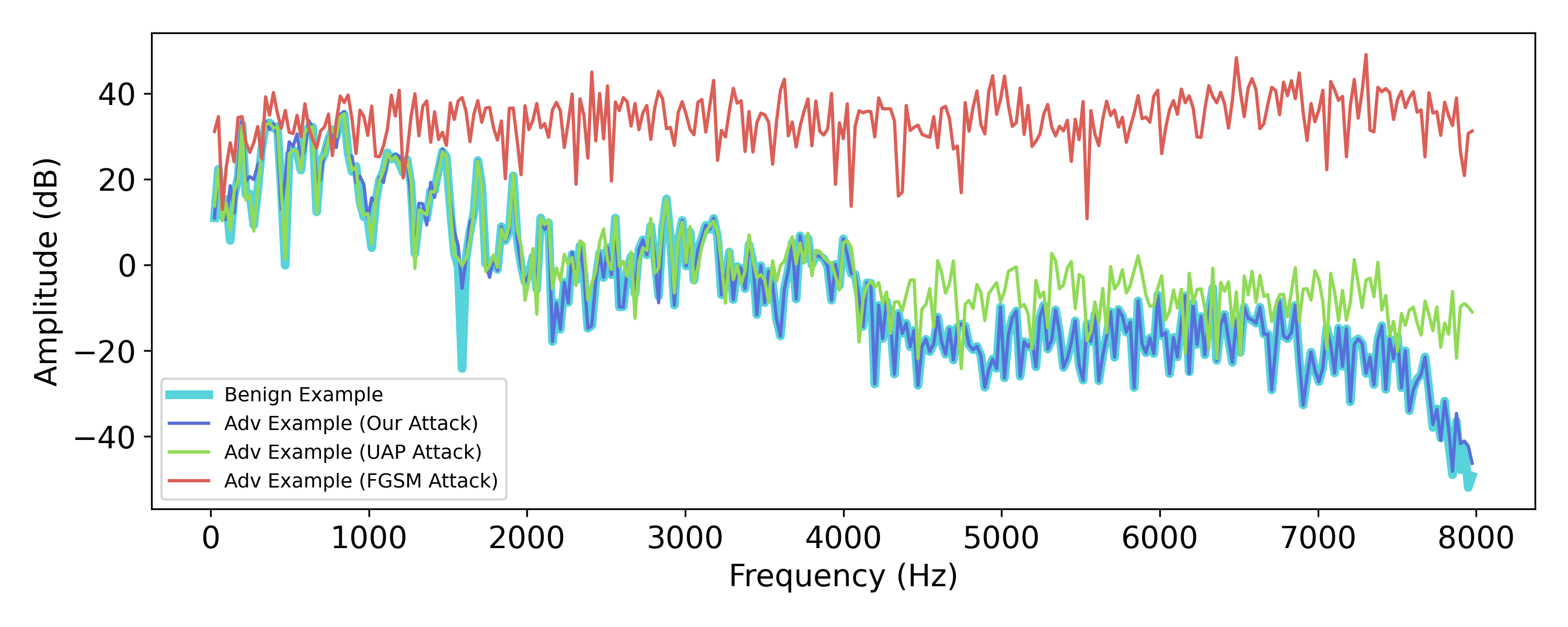}
   \caption{Frequency spectrum of sample learned adversarial examples, $x_i + v$, on the Speech Commands dataset. The amplitude of each learned attack has been adjusted so that the SNR is fixed at 10 dB. The frequency spectrum of our attack is much closer to the frequency spectrum of the benign example compared to the baseline UAP and FGSM attacks.}
   \label{fig:fourier_spectrum}
\end{figure}

Let $\mathcal{D} = \{ (x_i, y_i) \}_{i=0} ^n $ be a training dataset of time-series, $x_i \in \mathbb{R}^T$ sampled at rate $f_s$ with labels $y_i$.  Our proposed method learns a single attack, $v$ of length $T$. We refer to the attack in time space as $v_{\text{time}}$ and we refer to the attack in frequency space as $v_{\text{freq}}$. The attack can be converted between these representations using the fast Fourier transform (FFT) or the inverse fast Fourier transform (IFFT). Explicitly, $v_{\text{freq}} = \text{FFT}(v_{\text{time}})$ and $v_{\text{time}} = \text{IFFT}(v_{\text{freq}})$. Because our datasets are real-valued, we take $\text{Re}(v_{\text{time}})$ as the final universal attack.  

The proposed implementation of the attack is to repeatedly play the trained attack on a loop while data is recorded intermittently as depicted in Fig.~\ref{fig:looped_attack}. In order for the attack to be effective it must therefore be time invariant, meaning that the attack remains effective regardless of what point in its cycle the recording begins at. To ensure time invariance, we advance the attack by a random time shift of $t$ at each pass through the model during training. This is implemented in frequency space, by multiplying the attack's Fourier coefficient for frequency $k$, $v_{freq}[k]$, by $e^{i \frac{2 \pi k t}{T}}$. 

We also constrain the frequency spectrum of the attack to match the frequency spectrum of the original time series,  $\{ x_i \}_{i=0} ^n $, to ensure that the attack is not easily detectable or removed through filtering. Specifically, during the first phase of training, we require that each Fourier coefficient of $x_i + v$ be no more than twice the corresponding Fourier coefficient of $x_i$. We use the loss term, $\mathcal{L}_{\text{spectrum}_1}$, defined in (\ref{eq:spectrum loss_linear}), to enforce this constraint. Then, during the second phase of training we replace $\mathcal{L}_{\text{spectrum}_1}$ with  $\mathcal{L}_{\text{spectrum}_2}$, defined in  (\ref{eq:spectrum loss_dB}), which compares the Fourier spectrum on a log scale, rather than linear scale. This second phase of training accounts for the different scales of the Fourier coefficients and enables the attack to better match the frequency spectrum of the training dataset, even when the Fourier coefficients are small. 

\begin{equation}
\label{eq:spectrum loss_linear}
 \mathcal{L}_{\text{spectrum}_1} =  \sum_{i=0}^n{ \text{ReLU}(|\text{FFT}(x_i + v))| - 2 * |\text{FFT}(x_i)|)} 
\end{equation} 

\begin{equation}
\label{eq:spectrum loss_dB}
 \mathcal{L}_{\text{spectrum}_2} =  \sum_{i=0}^n{ \text{ReLU}(20 \text{log}_{10}(\frac{|\text{FFT}(x_i + v))|}{ 2 * |\text{FFT}(x_i)|}) }
\end{equation} 


In Fig. \ref{fig:fourier_spectrum}, we show the frequency spectrum of an adversarial example, $x_i + v$ trained on the Speech Commands dataset, as well as the frequency spectrum of the baseline UAP and FGSM attacks. The frequency spectrum of our attack is closely aligned with the frequency spectrum of the unperturbed benign example, whereas the other attacks have high frequency components not present in the original data. In section~\ref{sec:filtering}, we demonstrate that these other attacks are much more vulnerable to being removed through low-pass filtering than our attack is. 

The other loss term we train with, $\mathcal{L}_{\text{classifier}}$, is the negative cross-entropy loss on the model prediction, $f(x_i + v)$, to ensure that the adversarial attack fools the model. The full training procedure is outlined in detail in Algorithm \ref{alg:train}. Note that the classification model $f$ refers to the composition of the spectrogram prepossessing step and the classifier network.

\begin{algorithm}[t]
\caption{Training Procedure for Universal FFT Attack.}\label{alg:train}

\textbf{Input:} Training dataset $D = \{ (x_i, y_i) \}_{i=0}^n$, with time series $x_i \in \mathbb{R}^T$ sampled at rate $f_s$ with labels $y_i$, trained classification model $f$, desired number of training epochs $N$\\
\textbf{Output:} Attack vector $v \in  \mathbb{R}^T$
\begin{algorithmic}[1]
\State Initialize the vector in the frequency domain $v_{\text{freq}}
\leftarrow 0$
\While{Current epoch $ < N$}
\For{each $(x_i, y_i) \in \mathcal{D}$}
\State Sample a random time shift, $t \sim \text{Uniform}[0, \frac{T}{f_s} ]$, 
\For{$k$ in $[\frac{-f_s}{2} , ... , \frac{f_s}{2} - \frac{f_s}{T} ,\frac{f_s}{2}]$ }
\State $v_{\text{freq}}[k] = v_{\text{freq}}[k] e^{i \frac{2 \pi k t}{T}} $
\EndFor
\State Transform, $v_{\text{time}} = \text{Re}(IFFT(v_{\text{freq}}))$
\State $\hat{y} = f(x_i + v_{\text{time}})$
\If{Current epoch $ < 0.8 * N$}
    \State $ \mathcal{L} = \mathcal{L}_{\text{classifier}}(y, \hat{y}) + \beta
 \mathcal{L}_{\text{spectrum}_1}(x_i, x_i + v_{\text{time}})$ 
\Else
    \State $ \mathcal{L} = \mathcal{L}_{\text{classifier}}(y, \hat{y}) + \beta
 \mathcal{L}_{\text{spectrum}_2}(x_i, x_i + v_{\text{time}})$ 
\EndIf
 
 \State Update the perturbation, $v_{\text{freq}} \leftarrow v_{\text{freq}} + \alpha \nabla {v_{\text{freq}}}  \mathcal{L}$
\EndFor
\EndWhile
\State \textbf{return} $\text{Re}(v_{\text{time}})$, real-valued attack in the time domain
\end{algorithmic}
\end{algorithm}

\section{Results}

\begin{figure*}[t!]
    \centering
    \includegraphics[ scale=0.25]{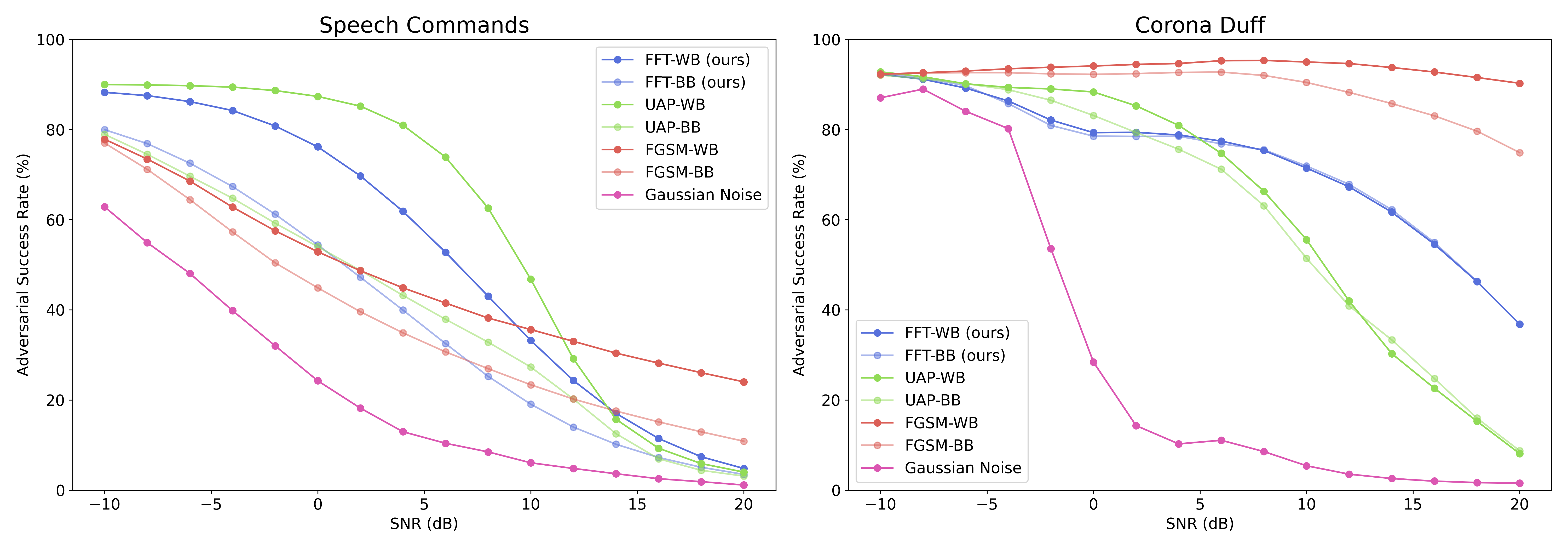}
   \caption{Adversarial success rate as a function of the SNR on the Speech Commands dataset (left) and on the Corona Duff dataset (right). We evaluate each attack on a white box (WB) model, as well as on a black box (BB) model.}
   \label{fig:asr}
\end{figure*}

\begin{figure*}[t!]
    \centering
    \includegraphics[ scale=0.25]{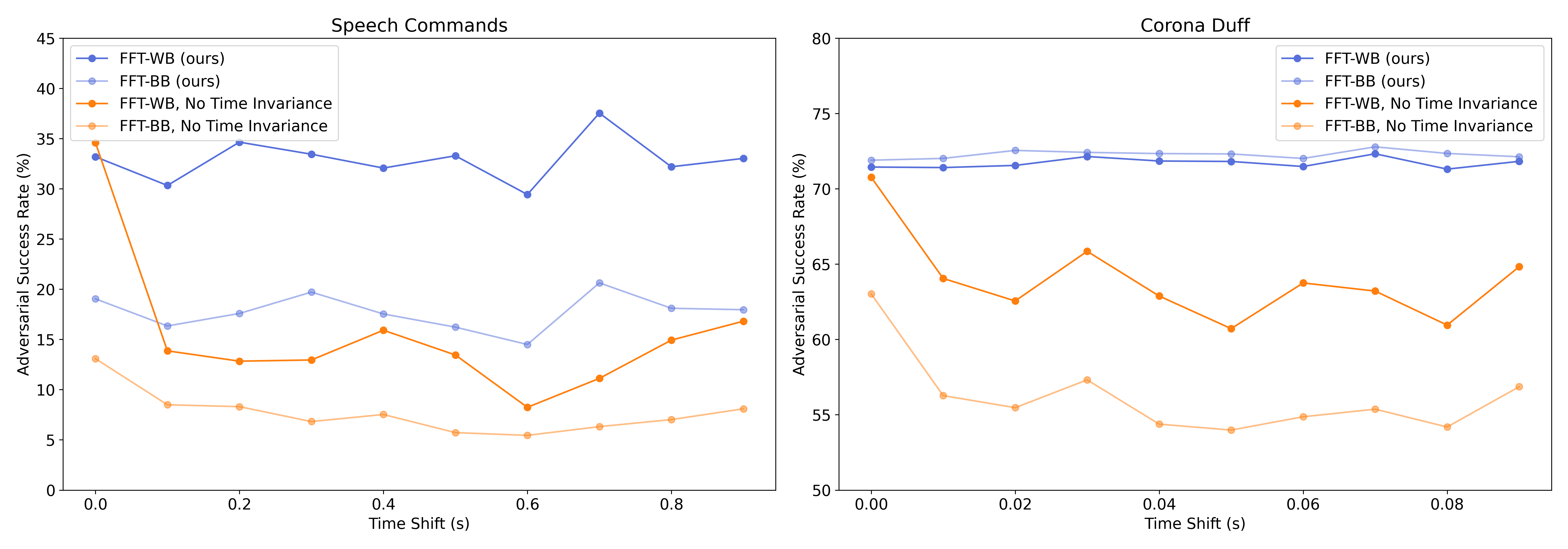}
   \caption{Adversarial success rate as a function of the time shift on the adversarial attack. At time shift $t$, the attack used is the segment of the looped attack from time $t$ to time $t+T$ where $T$ is the length of the input being attacked.}
   \label{fig:time-invariance}
\end{figure*}

\begin{figure*}[t!]
    \centering
    \includegraphics[ scale=0.25]{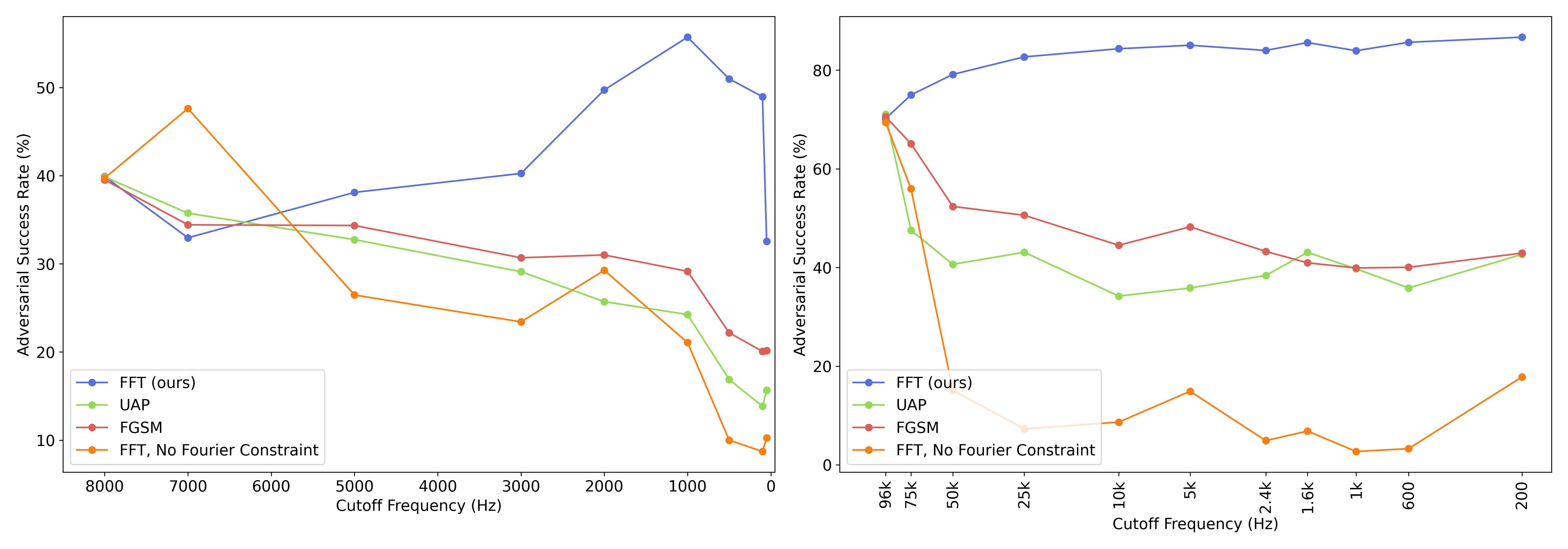}
   \caption{Adversarial success rate of each attack on a classifier that preprocesses inputs using a low-pass filter with the indicated cutoff frequency. The SNR for each attack has been adjusted so that the adversarial success rate at the highest cutoff frequency is comparable.  }
   \label{fig:freq_test}
\end{figure*}

We evaluate the ASR of the adversarial examples, $x + \alpha v$, where $v$ is the attack and $\alpha$ is adjusted to control the SNR. We evaluate the attacks on a white box (WB) model, the model the attack was trained on, as well as on a black box (BB) model. For black box evaluation, we assume that the attacker does not have access to the model weights, but does have access to the model architecture and the model training data. 

As depicted in Fig.~\ref{fig:asr}, on the Speech Commands dataset for all SNRs below 14 dB, the ASR of our attack on a white box model (FFT-WB) is above 20\%. We emphasize that our goal is not necessarily to create a state-of-the-art attack, but rather to create an attack with an ASR that is significant enough that a defender would not be able to ignore the attack and would likely spend resources against the attack. On a black box model, the ASR of our attack (FFT-BB) decreases, however the ASR is at least 20\% for SNRs up to 10 dB.

On the Corona Duff dataset, the ASR of that attack is at least 35\% for all SNRs tested. Additionally, on this dataset, the attack is almost as effective on a black box model as it is on a white box model. While FGSM does outperform our attack, particularly at high SNRs, we note that our attack is a universal attack, whereas FGSM must be calculated for each individual input. Visualizations of the learned attacks are available in Appendix~\ref{appendix:viz}. 


\subsection{Time Invariance}

In Fig.~\ref{fig:time-invariance}, we test the time invariance of the attack with the SNR fixed at 10 dB. We repeatedly play the attack on a cycle and shift the start time of the speech recording or URE data as depicted in Fig.~\ref{fig:looped_attack}. The plot also includes an ablation study, where we removed the time invariance transformation from the attack training by fixing the random training time shift to $t = 0$ in line 4 of Algorithm~\ref{alg:train}. 

Both our white box (FFT-WB) and black box (FFT-BB) attack have a relatively constant ASR on the Speech Commands dataset, ranging between about 30-37\% and 15-20\%, respectively. In contrast, the white box ablation attack has a high ASR of 35\% with a time shift of zero at evaluation time, but for all evaluation time shifts greater than zero the ASR drops below 15\%. The black box ablation attack is also sensitive to the evaluation time shift, dropping from 13\% to below 8\% for nonzero evaluation time shifts. The results with the Corona Duff dataset are similar as both the white box and black box versions of our attack have an ASR between 70\% and 73\% at all evaluation time shifts, and the ablation attack exhibits a decreased ASR when the evaluation time shift is nonzero. 

\subsection{Robustness to Filtering}
\label{sec:filtering}

To evaluate the robustness of the attack to filtering, we propose a set-up where a defender receives an input, which could be benign or adversarial.  The defender filters the received input and then evaluates it on a model that has been trained on filtered benign data. Since our datasets are composed of mostly low frequency information, we use low-pass filtering. If the received input is adversarial and the attack is composed of high frequencies, the filtering should remove most of the attack before the model makes its prediction, thus reducing the effectiveness of the attack. If the received input is benign, the model prediction should be accurate because the model has been trained on filtered benign data. 

More formally, let $f_k$ denote a classifier which has been trained on data low-pass filtered with cutoff frequency $k$. Then, $f_k(\textit{lowpass}_k(x) )$ is the classifier prediction on benign input $x$ and $f_k(\textit{lowpass}_k(x + \alpha v) )$ is the classifier prediction on adversarial input $x + \alpha v$. We evaluate the ASR of our attack and our baseline attacks on $f_k$ for a range of frequencies. Note that the attacks are the same attacks as in the previous sections. These were learned on an unfiltered model with unfiltered data and were not trained on the $f_k$ they are evaluated on.

Fig.~\ref{fig:freq_test} depicts the results.  For reference, we also include a version of our attack trained without either $ \mathcal{L}_{\text{spectrum}}$ loss term so that the frequency spectrum of this ablation attack is not constrained to match the frequency spectrum of the benign data. We adjust the SNR of each attack so that the ASR of each attack is comparable when the cutoff frequency is half the sampling rate.  

\begin{figure*}[h!]
    \centering
    \includegraphics[scale=0.25]{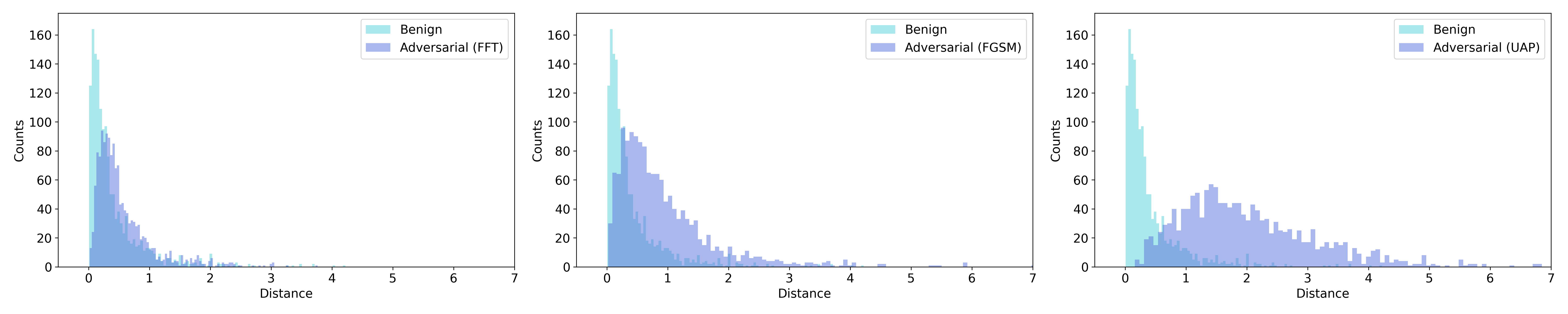}
   \caption{ $L_2$ distance distributions between classifier predictions of an input, which could be adversarial or benign, and the transformed input after MP3 compression on the Speech Commands dataset. From left to right the adversarial inputs are generated using our FFT attack, FGSM, and UAP.}
   \label{fig:defense_distances}
\end{figure*}
 
\begin{table*}[h!]
\caption{ AUC score of each transformation on each attack. All attacks are evaluated with SNR = 10 dB. The first three columns refer to attacks on the Speech Commands dataset (SC) and the last three columns refer to attacks on the Corona Duff (CD) dataset. }
\label{table:defenses}
  \centering
  \renewcommand{\arraystretch}{1.2}
\begin{tabular}{|r|r|r|r|r|r|r|}
    \hline \small{Transformation} & \small{FFT-SC}  & \small{UAP-SC}  & \small{FGSM-SC} & \small{FFT-CD}  & \small{UAP-CD}  & \small{FGSM-CD}
 \\ \hline
MP3 Compression & \textbf{0.60 ± 0.05} & 0.92 ± 0.13 & 0.68 ± 0.08  &\textbf{ 0.56 ± 0.03} & 0.84 ± 0.12 & 0.70 ± 0.14  \\ \hline
Quantization & \textbf{0.57 ± 0.03} & 0.74 ± 0.18 & 0.73 ± 0.18 & \textbf{0.54 ± 0.02} & 0.80 ± 0.07 & 0.88 ± 0.06   \\ \hline
Down-Up Sampling &\textbf{0.52 ± 0.01} & 0.89 ± 0.07 & 0.65 ± 0.05 &\textbf{ 0.54 ± 0.01} & 0.79 ± 0.12 & 0.71 ± 0.12   \\ \hline
Noise Flooding & \textbf{0.59 ± 0.05} & 0.76 ± 0.14 & 0.82 ± 0.04 & \textbf{0.58 ± 0.02} & 0.95 ± 0.06 & 0.95 ± 0.05  \\ \hline
\end{tabular}
\end{table*}

On the Speech Commands dataset with low pass filtering  the ASR of our attack is greater than or equal to 32\% for all frequencies tested. In contrast, for all other attacks, as lower cutoff frequencies are used, the effectiveness of the attack is reduced from an ASR of 40\% at a cutoff frequency of 8 kHz to below 20\% for cutoff frequencies of 500 Hz and below. As depicted in Fig.~\ref{fig:fourier_spectrum}, the UAP and FGSM attacks have high frequency components which low-pass filtering removes. 

Similarly, on the Corona Duff dataset our attack has an approximately constant ASR, averaging 82\% across all cutoff frequencies tested. In contrast, the other attacks demonstrate a significant reduction in ASR with filtering from a 70\% ASR at 192 kHz to below a 40\% ASR for the UAP and FSGM attacks. Thus, by restraining our attack to only use the low frequencies present in the original data, we have designed an attack that is much more robust to this filtering test than any of the baseline attacks tested. 

\subsection{Transform and Compare Defenses}

Besides filtering, to defend speech recognition systems against white box adversarial attacks, several works have proposed a tranform-and-compare pipeline to detect adversarial examples. Given an input, which could be benign or adversarial, this pipeline compares model predictions on the input and a transformed version of the input. Several different transformation functions have been proposed including the addition of random noise \cite{Rajaratnam_2018, dong2021}, audio compression \cite{Rajaratnam2018_b, zhang2019}, quantization \cite{hussain2021waveguard}, down-up sampling \cite{hussain2021waveguard},  and filtering \cite{kwon2020}.  If the distance between the model predictions on the transformed and original input is higher than a threshold, the input is flagged as adversarial because adversarial examples are generally less robust against perturbations than benign examples are. 

In the speech recognition system defense pipeline, the character error rate is typically used to measure the distance between model predictions. However, because our models are generic classifiers rather than speech-to-text models, we use the $L_2$ distance between the model outputs of the original input and transformed input. We calculate this distance for 800 benign inputs and 800 adversarial inputs. From the resulting distributions, a threshold can then be determined to use for flagging adversarial examples. 

Using MP3 compression as the transformation function on Speech Commands data, we plot the distribution of distances between the original and transformed benign inputs and the distribution of distances between the original and transformed adversarial inputs for our attack (FFT) and the FGSM and UAP baselines in Fig.~\ref{fig:defense_distances}. For the FGSM and UAP attacks, the distances for adversarial data are generally larger than for benign data, making adversarial examples generated with these attacks more easily identifiable. For our attack, we find much more overlap in the distribution of benign distances and the distribution of adversarial distances, demonstrating that our attack is much more difficult to flag using this defense method.

 Table ~\ref{table:defenses} reports the area under the curve (AUC) for all transformations tested, averaged over 5 runs of attack training. Full details of each transformation function, as well as the full AUC plots, are in Appendix ~\ref{appendix:defenses}. On both datasets, for all transformations tested, our attack has an AUC score close to $0.50$, averaging an AUC of $0.57$ across all transformations on Speech Commands and $0.56$ on Corona Duff. For comparison, the AUC scores of the baseline attacks are much higher with average AUC scores of $0.87$ and $0.81$ for the UAP and FGSM attack on Speech Commands and $0.83$ and $0.72$ for the UAP and FGSM attack on Corona Duff. This indicates that our attack is more difficult to identify  using the tranform-and-compare defense pipeline.

\subsection{Transferability to Speech-to-Text Model}

\begin{table*}[h!]
\caption{Transferability of attacks trained on our speech classifier to the non-classification task of speech-to-text translation. We use the Deep Speech speech-to-text model and report the model accuracy (ACC), character error rate (CER), and adversarial success rate (ASR) at a fixed SNR of 5 dB. }
\label{table:deepspeech}
  \centering
\begin{tabular}{|r|r|r|r|}
    \hline  & ACC & CER & ASR
 \\ \hline
    No Attack & 64.30  & 23.74  & 0.00  \\ \hline
    FFT   & 34.72 ± 4.7 & 51.52 ± 6.2 & 50.52 ± 6.51  \\ \hline
    FFT (${L}_{\text{spectrum}_1}$ only)  & \textbf{18.47 ± 4.09} & \textbf{73.52 ± 9.92} & \textbf{73.64 ± 5.61}   \\ \hline
    FGSM  & 26.24 ± 2.08 & 58.15 ± 2.47 & 62.00 ± 2.87  \\ \hline
    UAP  & 28.91 ± 1.62 & 55.56 ± 2.04 & 58.73 ± 2.25   \\ \hline
    Gaussian Noise & 36.64 ± 0.65 & 47.41 ± 0.77 & 48.25 ± 0.79  \\ \hline

\end{tabular}
\end{table*}

The universal attack trained on the Speech Commands classifier also transfers to a speech-to-text model. We test the learned attack on a pretrained, off-the-shelf Deep Speech \cite{name} model. In Table~\ref{table:deepspeech} we report the accuracy (ACC), the Character Error Rate (CER) and ASR of our attack with a fixed SNR of 5 dB. The CER is defined as $(S_c + D_c + I_c)/N_c$ where $S_c, D_c,$ and $I_c$ refer to the number of character substitutions, deletions, and insertions, respectively. 

While our attack as proposed achieves metrics only slighty better than just adding random Gaussian noise to the benign data, if we only use ${L}_{\text{spectrum}_1}$ in training and forego the second training phase with ${L}_{\text{spectrum}_2}$, the attack is much more effective and transfers very well. 



\section{Conclusion}
We presented an adversarial attack for general time series data designed for real-world implementation. We demonstrate that for both speech and URE data, the universal attack is time invariant, robust to filtering, and is robust to common transform-and-compare defense pipelines. In the future, it would be interesting to test the attack in a real-world environment as this initial design and testing of the attack suggest that the attack may be well-suited to real-world implementation. 




\bibliography{references}
\bibliographystyle{IEEEtranS}

\newpage
\onecolumn
\appendix
\section{Appendix}

\subsection{Attack Visualizations}
\label{appendix:viz}

\begin{figure}[h!]
    \centering
    \includegraphics[scale=0.45]{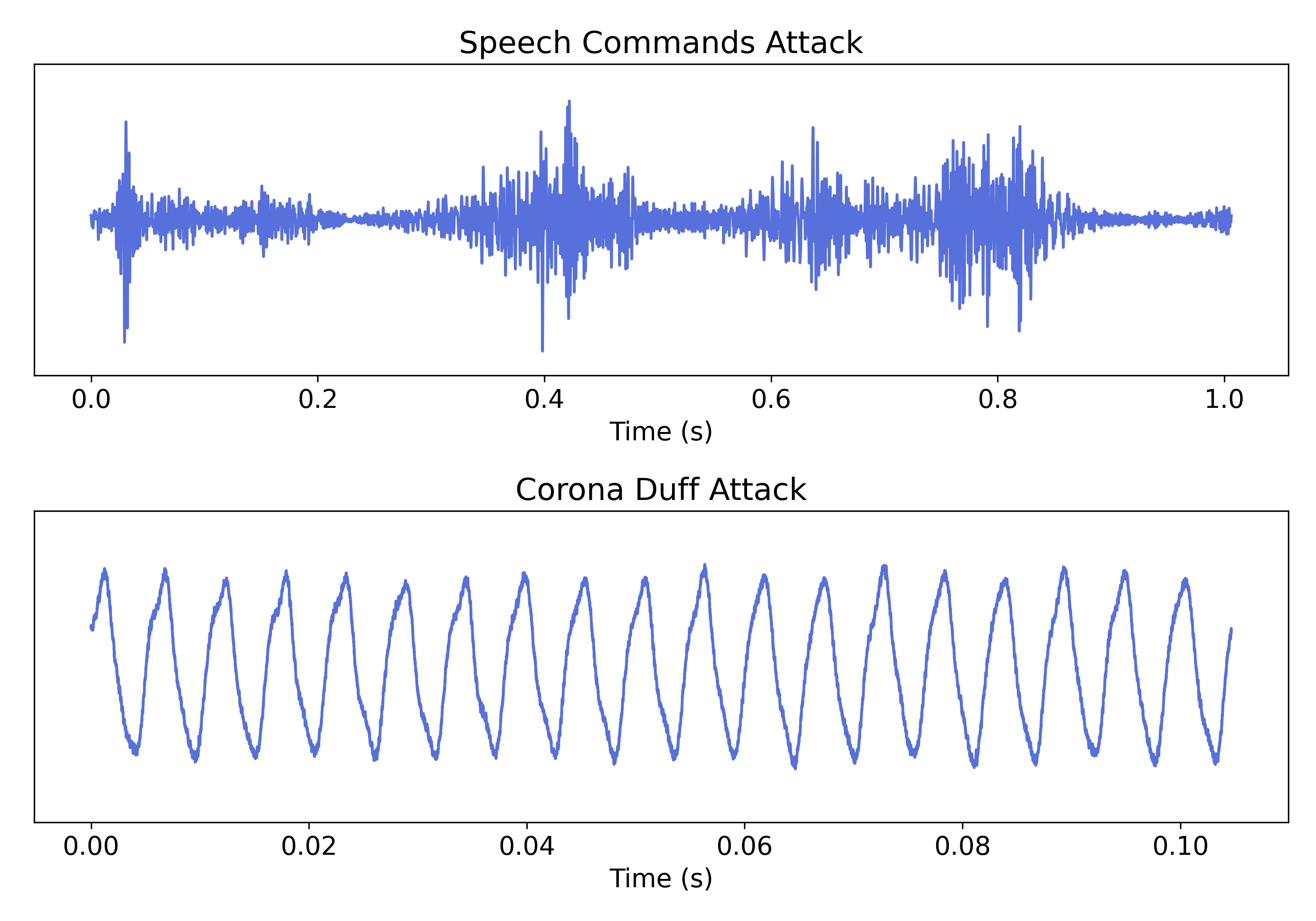}
   \caption{Our learned universal attack on each dataset, in time space.}
   \label{fig:learned_attacks}
\end{figure}

\begin{figure}[h!]
    \centering
    \includegraphics[scale=0.45]{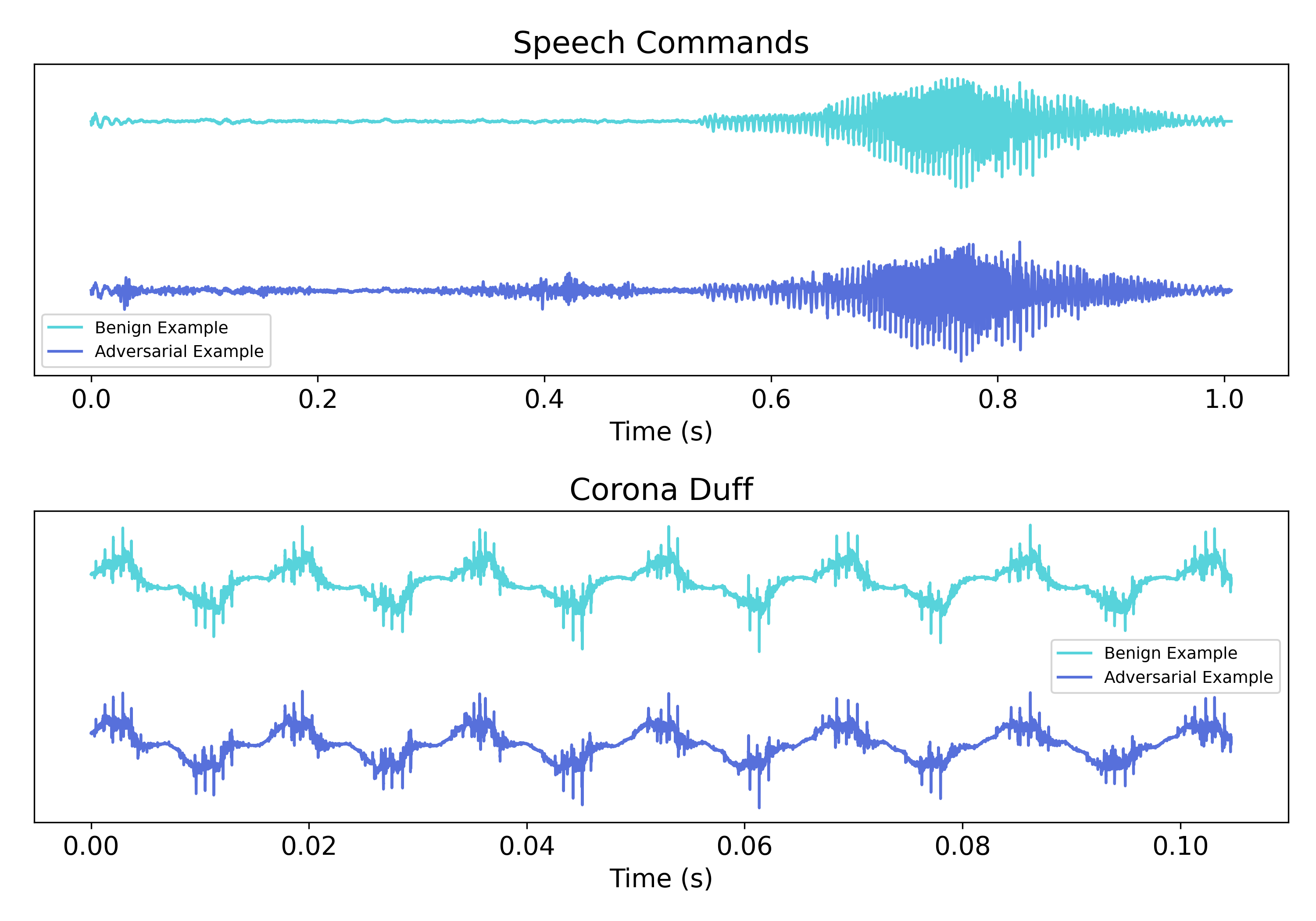}
   \caption{Sample adversarial examples on each dataset in time space. The lighter colored time series is the original, unperturbed benign example and the darker colored time series is the adversarial example with an SNR of 10 dB. }
   \label{fig:learned_examples}
\end{figure}

\subsection{Defenses}
\label{appendix:defenses}
Here we provide full details of the transformation functions tested in the transform-and-compare defense pipeline. The quantization function quantizes inputs to an 8-bit representation and then dequantizes inputs. We use the TensorFlow implementation. The down-up sampling function downsamples inputs to half the original sample rate and then upsamples this sequence back to the original sample frequency. With noise flooding Gaussian noise with standard deviation $0.01$ is added to the inputs. We additionally tested noise flooding specific frequency bands by filtering the Gaussian noise with a band-pass filter as in ~\cite{Rajaratnam_2018}. The results for noise flooding without filtering are very similar to noise flooding with band-pass filtering so we just report the scores for noise flooding without filtering. 

In Fig.~\ref{fig:roc_curves}, we plot ROC curves or each transformation used. For both datasets and all transformations used, our attack is the least detectable under the transform-and-compare pipeline. 

\begin{figure*}[t!]
    \centering
    \includegraphics[ scale=0.25]{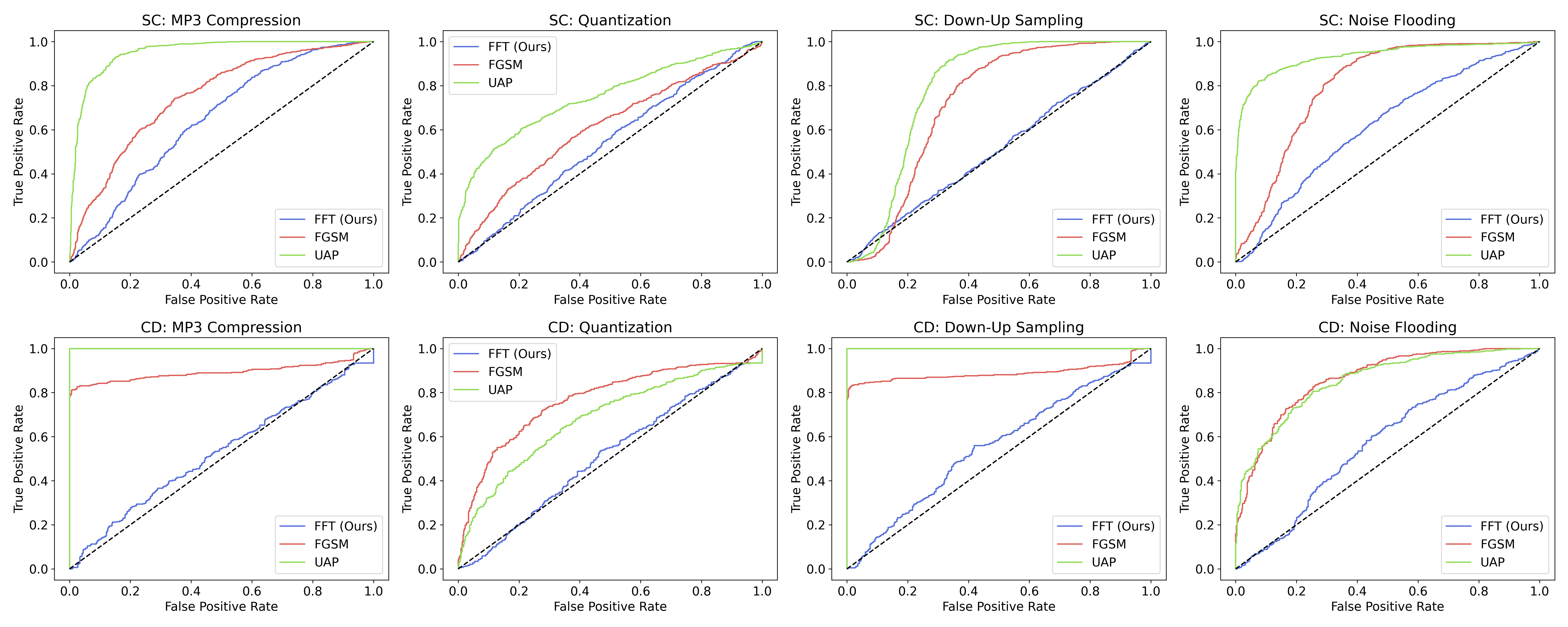}
   \caption{ROC curves for each transformation used in the defense transform-and-compare pipeline. The top row depicts results for the Speech Commands (SC) dataset, and the bottom row depicts results for the Corona Duff (CD) dataset.}
   \label{fig:roc_curves}
\end{figure*}


\end{document}